\documentstyle[preprint,aps]{revtex}
\tighten

\title{\bf Vortex ordering in fully-frustrated \\
superconducting systems with dice lattice}
\author{S. E. Korshunov}
\address{L. D. Landau Institute for Theoretical Physics, Kosygina 2, Moscow 117940, Russia}
\date{26 July 2000}
\draft
\begin{document}
\maketitle

\begin{abstract}
The structure and the degeneracy of the ground state of a
fully-frustrated $XY$-model are investigated for the case of a
dice lattice geometry. The results are applicable for the
description of Josephson junction arrays and thin superconducting
wire networks in the external magnetic field providing
half-integer number of flux quanta per plaquette. The mechanisms
of disordering of vortex pattern in such systems are briefly
discussed.

\end{abstract}

\section{Introduction}

De Gennes \cite{dG} and Alexander \cite{A} have shown that the
linearized Ginzburg-Landau equations for a superconducting wire
network in external magnetic field can be mapped on the eigenvalue
equations for a single electron hopping problem in the same
geometry. Thus the dependence of a (mean-field) superconducting
transition temperature on external field can be found by following
the field dependence of the lowest eigen-value
in single electron problem.

Recently it has been shown by Vidal {\em et al} \cite{VMD} that
the single electron hopping problem has very special features in
the case of so-called \cite{HC,Su} dice lattice (Fig. 1) if the
value of the magnetic flux per plaquette $\Phi$ is equal to one
half of the  flux quantum $\Phi_0$. Namely, the spectrum of an
electron lacks any dispersion and is reduced to three discrete
levels.

This theoretical result has led to the interest in experimental
investigation of superconducting networks with dice lattice
geometry \cite{ABF,PAS}. It has been shown \cite{PAS} that in
contrast to the case of $\Phi=\Phi_0/3$, for which  vortex pattern
in dice  network is nicely ordered, at $\Phi=\Phi_0/2$ [so-called
fully-frustrated (FF) case] the vortices do not form (presumably
at the same temperature) any regular pattern. The authors of Ref.
\cite{PAS} have suggested that this absence of ordering is related
with an infinite degeneracy and localized structure of the states
corresponding to the lowest energy level in terms of a single
electron problem (or to the lowest free energy in terms of a
superconducting network).

Although this conjecture can be correct, it still has to be verified.
A superconducting network problem reduces to
a single electron hopping problem only at the mean-field transition
point. Below it the non-linear terms in Ginzburg-Landau equation
become important and may completely or partially remove the high
degeneracy of the state with the lowest free energy.

In the present work we use    a different approach for theoretical
investigation of an ordering in a FF superconducting  system
with dice lattice.
We consider another limit when the amplitude of the order
parameter is well defined and uniform, but phase fluctuations are
possible and can lead to destruction of an ordered state.
In that limit a discrete superconducting system
(a wire network or a junction array)
in external magnetic field can be described
by a frustrated $XY$-model introduced in Sec. 2.

In Sec. 3 we propose a highly symmetric state which due to
simplicity of its structure may be a good candidate for the ground
state of a FF $XY$-model with dice lattice. In Sec. 4 we show that
this state has high additional degeneracy because it allows for
formation of zero-energy domain walls. Sec. 5 is devoted to a
brief discussion of possible consequences of this additional
degeneracy for the disordering of vortex pattern in FF
superconducting systems with dice lattice geometry.

\section{The model}

In the regime when only phase fluctuations are of importance
an array of weakly coupled superconducting islands
can be described by the Hamiltonian:
\begin{equation}
H=\sum_{\langle ij \rangle}^{}V(\theta_{ij}),    \label{a1}
\end{equation}
where the sum is performed over all pairs of coupled islands and
\begin{equation}
\theta_{ij}=\varphi_j-\varphi_i-\frac{2\pi}{\Phi_0}
\int_{i}^{j}d{\bf x}\;{\bf A}({\bf x})\equiv-\theta_{j\,i}  \label{a2}
\end{equation}
is the gauge-invariant phase difference which can be associated
with the link $\langle ij\rangle$.
Here $\varphi_j$ is the order parameter phase of $j$-th
superconducting island and ${\bf A}$ is the vector potential.
The phases $\varphi_j$ are defined up to a shift by a multiple
of $2\pi$, therefore the interaction function $V(\theta_{})$
has to be periodic in $\theta$.
The form of $V(\theta{})$ depends on the type
of the coupling. For Josephson junction array
\begin{equation}
V(\theta_{})=-J\cos\theta_{},                  \label{a3}
\end{equation}
where $J$  is the coupling constant of the junction.

Summation of Eq. (\ref{a2}) over a perimeter of a lattice
plaquette imposes a constraint
\begin{equation}
\sum_{\Box}^{}\theta_{ij}=-2\pi f,
                                                \label{a5}
\end{equation}
where the frustration parameter $f$ is equal to $\Phi/\Phi_0$ and
$\Phi$ is the magnetic flux
threading the plaquette. In the limit when screening effects can be
neglected $\Phi$ is determined by the external field and for the
uniform field and flat geometry is proportional to the area of the
plaquette. If all the plaquettes have equal areas
the value of $f$ is the same for all plaquettes.
In such case a system is called uniformly frustrated.

When interaction function $V(\theta)$ is periodic in $\theta$ it
is convenient to consider variables $\theta_{ij}$ reduced to the
interval $[-\pi,\pi]$. That makes the description of any state in
terms of $\theta_{ij}$ more transparent, but transforms the
constraint (\ref{a5}) into
\begin{equation}
\sum_{\Box}^{}\theta_{ij}=2\pi M,~~M\equiv m-f;
                                                \label{a6}
\end{equation}
where $m$ is an integer. The form of Eq. (\ref{a6}) shows
that $f$ can be reduced to the interval $-1/2< f\leq 1/2$,
all other values of $f$ being equivalent to some value from that
interval.
If $V(\theta{})$ is an even function $f$ is additionally
equivalent to $-f$.
The case of $f=1/2$ is usually called a fully-frustrated (FF) $XY$-model.

Well below mean-field transition temperature a network of thin
superconducting wires can be described by the same Hamiltonian
(\ref{a1}) with the term $\varphi_j-\varphi_i$ in the definition
of $\theta_{ij}$ [Eq. (\ref{a2})] now substituted by the integral
$\int_{i}^{j}dx(d\varphi/dx)$ along the link $\langle ij\rangle$
\cite{AH}. In that case summation of Eq. (\ref{a2}) around a
perimeter of a plaquette leads directly to Eq. (\ref{a6}). In the
limit of long thin wires the interaction function is almost
harmonic \cite{AH}:
\begin{equation}
V(\theta_{})\propto \theta_{}^2.        \label{a7}
\end{equation}

In this work we use the term "frustrated $XY$-model" for a system
defined by Eqs. (\ref{a1}) and (\ref{a6}) with a general form of
the interaction function and not only for $V(\theta)$ of the form
(\ref{a3}). Thus our approach is valid for the description both of
junction arrays and wire networks.

\section{The ground state}

In a FF $XY$-model all variables $M$ are half-integer and
different low-lying  extrema of the Hamiltonian can be
characterized by the distribution of positive and negative
half-vortices ($M=\pm 1/2$) in the plaquettes of the lattice. The
vortices of the same sign repeal each other, therefore in the
ground state they can be expected to be situated as far from each
other as possible. In particular, in the case of a FF $XY$-model
with square lattice the positive and negative half-vortices form
in the ground state a regular checkerboard pattern \cite{V}.
Analogous pattern in which the nearest neighbors of each
half-vortex are of the opposite sign is possible when they occupy
the sites of a honeycomb lattice, that is in FF $XY$-model with
triangular lattice \cite{MS,LCJW}.

A dice lattice is dual to a Kagom\'{e} lattice, therefore in a FF
$XY$-model with dice lattice the half-vortices can be considered
as occupying the sites of a Kagom\'{e} lattice. Since a Kagom\'{e}
lattice is constructed from triangles it is impossible to
distribute the half-vortices in it in such a way that all nearest
neighbors are of the opposite sign. The half-vortices of the same
sign will have to form clusters and the minimal size of such
clusters which allow for covering of a Kagom\'{e} lattice turns
out to be equal to three. The most symmetric example of a regular
(periodic) arrangement of vortices on a Kagom\'{e} lattice in
which half-vortices of the same sign form the clusters of the size
three (triads) is shown in Fig. 2a.

This state has the 12-fold degeneracy and can be described as a
regular lattice of vacancies (absent positive vortices) on the
background of $f=2/3$ ground state or, equivalently, a regular lattice
of extra positive vortices on the background of $f=1/3$ ground state
(cf. with Ref. \cite{PAS}).
However symmetry considerations show that its structure
in terms of gauge-invariant phase variables $\theta_{ij}$
(which is shown in Fig. 3a) is very simple
and can be constructed by repetition (with rotation and reflection)
of a simple three-link pattern shown in Fig. 3b.
Although we can not rigorously prove that this state has the lowest
possible energy (which is typical for $XY$-models with nontrivial
frustration), we believe that
the simplicity of its structure strongly supports this conjecture.

In the state depicted in Fig. 3a the variables $\theta_{ij}$
acquire only three different values which we denote $\theta_{a}$
($a=1,2,3$; $0<\theta_{1} <  \theta_{2} < \theta_{3}<\pi$) and
show in figures as single, double and triple arrows. The
half-vortices of the same sign are separated by single arrows, the
central half-vortex of each triad is separated from its neighbors
of the opposite sign by triple arrows and the lateral
half-vortices of opposite sign are separated by double arrows. The
same set of rules for extracting the distribution of $\theta_{ij}$
from the distribution of half-vortices  applies also to all the
other states with the same energy discussed below.

The energy of the considered state (calculated per triple site of
dice lattice) is given by
\begin{equation}
E=V(\theta_{1})+V(\theta_{2})+V(\theta_{3}),    \label{b1}
\end{equation}
whereas general constraints (\ref{a6}) are reduced to
\begin{equation}
2\theta_{1}+2\theta_{3} =  \pi \label{b2}
\end{equation}
for the central half-vortex of each cluster and
\begin{equation}
-\theta_{1}+2\theta_{2}+\theta_{3} = \pi \label{b3}
\end{equation}
for all the other (lateral) half-vortices. Variation of Eq.
(\ref{b1}) with constraints (\ref{b2})-(\ref{b3}) gives
\begin{equation}
  V'(\theta_{1})+V'(\theta_{2})=V'(\theta_{3}),    \label{b4}
\end{equation}
which (not unexpectedly) coincides with the condition of the current
conservation for each of the triple sites.
The current conservation at each of the six-link sites
in the state of Fig. 3a is ensured automatically (by symmetry).

For $V(\theta_{})$ of the form (\ref{a3}) (corresponding to Josephson
junction array) the solution of Eqs. (\ref{b2})-(\ref{b4}) gives
\begin{equation}
\theta_{1}=\arctan\frac{\sqrt{2}-1}{\sqrt{2}+1}\approx 10^\circ
,~~ \theta_{2}=\frac{\pi}{4}+\theta_{1}\approx 55^\circ,~~
\theta_{3}=\frac{\pi}{2}-\theta_1\approx 80^\circ;  \label{b5}
\end{equation} whereas for $V(\theta_{})\propto \theta_{}^2$
(the case of a thin wire network)
\begin{equation}
\theta_{1}=15^\circ,~~ \theta_{2}= 60^\circ,~~ \theta_{3}=
75^\circ.
                                                    \label{b6}
\end{equation}
Thus the values of $\theta_{a}$ are only weakly sensitive to the
type of superconducting system which manifests itself in the form of
the current-phase relation.

\section{The additional degeneracy}

The regular state depicted in Fig. 3a (Fig. 2a) allows for the
construction of zero energy domain wall. Fig. 3c shows how one can
rearrange the arrows in the lower half of Fig. 3a without
invalidating constraints (\ref{a6}) or current conservation
relations, obtaining in such way another extremum with the same
energy. The same state is shown in Fig. 2b in terms of
distribution of half-vortices. It looks like a domain wall
separating the state (a) ({\em i.e.} the state shown in Fig. 2a)
from another version of the same state in which the triads of
negative half-vortices change their orientation by $60^\circ$.

It is possible to construct such domain wall on each horizontal
line similar to the line shown in Fig. 2b. This increases the
degeneracy by the factor $2^{N}$ (where $N$ is the number of
available positions of the domain walls) due to the binary
possibility of having or not having a domain wall at each
available position. The  regular state constructed by inserting
into the state of Fig. 2a a domain wall at each available position
is shown in Fig. 2c. This state is also periodic, but has the
higher (24-fold) degeneracy than the state (a).

Insofar we have discussed only the zero-energy domain walls which
are parallel to the triads of positive half-vortices. It follows
from the symmetry considerations that analogous domain walls can
be also constructed in parallel to the triads of negative
half-vortices. However, the energy remains the same only if all
domain walls of this type have the same orientation, therefore the
total increase of the degeneracy due to a possible creation of
zero-energy domain walls of the type (b) is given by a factor
$2^{N+1}$. Analogous restriction for the creation of zero-energy
domain walls appears in the case of the frustrated $XY$-model with
a triangular lattice  and $f=1/4$ or $f=1/3$ \cite{KVB}.

In Fig. 3 the central half-vortex of each triad is marked by
square brackets. It is not hard to notice that all the other
(lateral) half-vortices form the rows of alternating pluses and
minuses. These rows are straight for the regular state of Fig. 3a
(Fig. 2a), but the presence of domain walls of the type (b)
makes them bend (in parallel to each other).

It turns out possible to interchange pluses and minuses in any of
these rows by interchanging the single and triple arrows on the links
which separate the row from the neighboring central half-vortices and
reversing the double arrows on all the links inside the row.
This procedure does not invalidate current conservation at any site
and does not change the energy of the system.

Such sign reversal in the rows of alternating lateral
half-vortices allows to construct the zero-energy domain wall of
different type shown in Fig. 2d which [like the domain wall of the
type (b)] also separates two different versions of the state (a),
but now with interchanged orientations of positive and negative
triads. A regular repetition of such domain wall at each available
position leads to the periodic state shown in Fig. 2e, which like
the state (a) has the 12-fold degeneracy.

The zero-energy domain walls of different types can cross each
other (as is shown in Fig. 2f) without increasing the energy of
the system. Each time time the wall of type (d) crosses the wall
of the type (b) it has to change its orientation by $60^\circ$,
all the walls of the type (d) being parallel to each other in each
strip between the walls of the type (b).

Thus the system simultaneously with an arbitrary number of the
walls of the type (b) can contain also an arbitrary number of the
walls of type (d). The total additional degeneracy [in comparison
with our reference state (a)] is given by the  factor $2^{N+M+1}$,
where $M$ is the number of positions available for domain walls of
the type (d), that is the number of rows of the alternating
lateral half-vortices. The most dense network of zero-energy
domain walls of both types produces the periodic state shown in
Fig. 2g, which like the state (c) is characterized  by the 24-fold
degeneracy.

\section{Discussion}

The zero-energy domain walls of the two types described above
produce the contribution to residual (zero-temperature) entropy
which grows with the increase of the system as $(N+M)\ln 2$, that
is slower than its area (which is proportional to $NM$). I. e.
they do not lead to appearance of the extensive residual entropy
in contrast to the case of the antiferromagnetic $XY$-model with a
Kagom\'{e} lattice in which the manifold of the ground states
allows for construction of zero-energy domain walls which can form
independent closed loops of arbitrary length \cite{HR,CCR}. This
property of the antiferromagnetic $XY$-model with Kagom\'{e}
lattice leads both to a finite extensive entropy \cite{HR,CCR} and
to the presence of a hierarchical sequence of barriers \cite{CCR}
which may explain the experimentally observed glass-like dynamics
of the antiferromagnet with such structure \cite{WDV}. The family
of the states considered in this work demonstrates less developed
degeneracy (analogous to that encountered in $f=1/3$ $XY$-model
with triangular lattice \cite{KVB}).

Since different values of $\theta_a$ correspond to different
values of $V''(\theta)$, at finite temperatures the accidental
degeneracy related to possible formation of zero-energy domain
walls can be expected to be removed due to the difference in free
energy of the small amplitude continuous fluctuations (spin waves)
in the same way as it happens in the $XY$-model with triangular
lattice and $f=1/4$ or $f=1/3$ \cite{KVB}. Most probably the spin
wave contribution to free energy will be minimal for the one of
the periodic states shown in Fig. 2, which therefore will be
dominant in the low temperature limit.

The zero energy domain walls separating from each other the
different versions of this state will then acquire a positive free
energy, so the phase transition associated with their
proliferation (and vortex pattern disordering) can be expected to
happen only at finite temperatures. However in the thin wire
networks with almost harmonic interaction function $V(\theta{})$
the effects related with the differences in spin wave free energy
will be extremely weak and therefore the disordering of vortex
pattern due to proliferation of domain walls may happen already at
rather low temperatures.

The strong disordering of vortex pattern observed in FF
superconducting network with dice lattice geometry \cite{PAS} can
also have some relation to geometrical irregularities. Gupta and
Teitel \cite{GT} have recently shown that in the case of the FF
$XY$-model with square lattice the irregularities of so-called
"positional disorder" type (uncorrelated lattice sites
displacements, etc.) produce an effective random field for the
Ising-type variables $M$ (describing the signs of the
half-vortices) and therefore induce the destruction of long-range
order (at large enough scales) even if the disorder is small. In
the system which allows for formation of the zero-energy domain
walls the relevance of this mechanism may be strongly amplified.

\section*{Acknowledgements}

This work was supported in part by the Program "Scientific Schools
of the Russian Federation" (grant No. 00-15-96747) and by the
Swiss National Science Foundation. The author is greatful to P.
Martinoli and B. Pannetier for introduction to the problem and
discussion of the results.

\newpage
\newcounter{nnx}
\newcounter{nny}
\newcommand{\stc}[2]{\setcounter{nnx}{#1}\setcounter{nny}{#2}}

\begin{figure}
\vspace{5mm}
\begin{center}
\setlength{\unitlength}{0.3mm}
\begin{picture}(200,160)(-10,0)
\thicklines \multiput(0,40)(60,0){4}{\line(0,1){80}}
\multiput(30,0)(60,0){3}{\line(0,1){60}}
\multiput(30,100)(60,0){3}{\line(0,1){60}}
\multiput(0,0)(60,0){3}{\line(3,2){60}}
\multiput(30,60)(60,0){2}{\line(3,2){60}}
\multiput(0,120)(60,0){3}{\line(3,2){60}}
\multiput(-10,72.5)(160,-12.5){2}{\line(3,2){40}}
\multiput(0,40)(60,0){3}{\line(3,-2){60}}
\multiput(30,100)(60,0){2}{\line(3,-2){60}}
\multiput(0,160)(60,0){3}{\line(3,-2){60}}
\multiput(-10,87.5)(160,12.5){2}{\line(3,-2){40}}
\end{picture}
\end{center}

\caption{Dice lattice is periodic and has hexagonal symmetry. It
consists of the sites with coordination numbers 3 and 6. All
elementary plaquettes are rhombic.\\}
\end{figure}
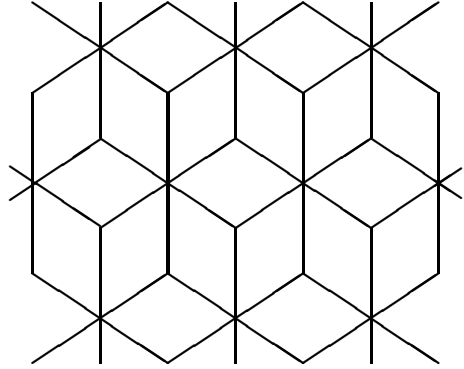

\newcommand{\appp}{\circle*{11}}
\newcommand{\ammm}{\circle{11}}
\newcommand{\putl}[1]{
\addtocounter{nnx}{+115}\addtocounter{nny}{-40}
\put(\value{nnx},\value{nny}){\makebox(0,0)[cc]{#1}}
\addtocounter{nnx}{-115}\addtocounter{nny}{+40}}
\newcommand{\mptp}[1]{\multiput(\value{nnx},\value{nny})(80,0){3}{\appp}\addtocounter{nnx}{#1}}
\newcommand{\mptm}[1]{\multiput(\value{nnx},\value{nny})(80,0){3}{\ammm}\addtocounter{nnx}{#1}}
\newcommand{\putp}[1]{\put(\value{nnx},\value{nny}){\appp}\addtocounter{nnx}{#1}}
\newcommand{\putm}[1]{\put(\value{nnx},\value{nny}){\ammm}\addtocounter{nnx}{#1}}

\setlength{\unitlength}{0.27mm}
\begin{center}
\begin{picture}(530,450)(0,-250)
\stc{0}{30} \putl{a)}

\mptp{20}\mptp{20}\mptp{20}\mptm{-50}\addtocounter{nny}{17}
\mptm{40}\mptm{-50}                  \addtocounter{nny}{17}
\mptm{20}\mptp{20}\mptp{20}\mptp{-30}\addtocounter{nny}{17}
\mptm{40}\mptm{-70}                  \addtocounter{nny}{17}
\mptp{20}\mptm{20}\mptp{20}\mptp{-50}\addtocounter{nny}{17}
\mptm{40}\mptm{-50}                  \addtocounter{nny}{17}
\mptp{20}\mptp{20}\mptm{20}\mptp{-30}\addtocounter{nny}{17}
\mptm{40}\mptm{-70}                  \addtocounter{nny}{17}
\mptp{20}\mptp{20}\mptp{20}\mptm{-50}\addtocounter{nny}{17}
\mptm{40}\mptm{-50}                  \addtocounter{nny}{17}
\mptm{20}\mptp{20}\mptp{20}\mptp{-30}\addtocounter{nny}{17}

\stc{300}{30} \putl{b)}

\mptp{20}\mptp{20}\mptm{20}\mptp{-50}\addtocounter{nny}{17}
\mptm{40}\mptm{-50}                  \addtocounter{nny}{17}
\mptp{20}\mptm{20}\mptp{20}\mptp{-30}\addtocounter{nny}{17}
\mptm{40}\mptm{-70}                  \addtocounter{nny}{17}
\mptm{20}\mptp{20}\mptp{20}\mptp{-50}\addtocounter{nny}{17}
\addtocounter{nnx}{-32}
\multiput(\value{nnx},\value{nny})(40,0){7}{\line(1,0){24}}
\addtocounter{nnx}{+32} \mptm{40}\mptm{-50} \addtocounter{nny}{17}
\mptp{20}\mptp{20}\mptm{20}\mptp{-30}\addtocounter{nny}{17}
\mptm{40}\mptm{-70}                  \addtocounter{nny}{17}
\mptp{20}\mptp{20}\mptp{20}\mptm{-50}\addtocounter{nny}{17}
\mptm{40}\mptm{-50}                  \addtocounter{nny}{17}
\mptm{20}\mptp{20}\mptp{20}\mptp{-30}\addtocounter{nny}{17}

\stc{0}{-220} \putl{c)}

\mptp{20}\mptp{20}\mptm{20}\mptp{-50}\addtocounter{nny}{17}
\mptm{40}\mptm{-50}                  \addtocounter{nny}{17}
\mptm{20}\mptp{20}\mptp{20}\mptp{-30}\addtocounter{nny}{17}
\mptm{40}\mptm{-70}                  \addtocounter{nny}{17}
\mptp{20}\mptp{20}\mptm{20}\mptp{-50}\addtocounter{nny}{17}
\mptm{40}\mptm{-50}                  \addtocounter{nny}{17}
\mptm{20}\mptp{20}\mptp{20}\mptp{-30}\addtocounter{nny}{17}
\mptm{40}\mptm{-70}                  \addtocounter{nny}{17}
\mptp{20}\mptp{20}\mptm{20}\mptp{-50}\addtocounter{nny}{17}
\mptm{40}\mptm{-50}                  \addtocounter{nny}{17}
\mptm{20}\mptp{20}\mptp{20}\mptp{-30}\addtocounter{nny}{17}

\stc{300}{-220} \putl{d)}
\addtocounter{nnx}{+45}\addtocounter{nny}{-25}
\multiput(\value{nnx},\value{nny})(20,34){7}{\line(3,5){10}}
\addtocounter{nnx}{-45}\addtocounter{nny}{+25}

\putp{20}\putp{20}\putp{20}\putm{20}
\putm{20}\putp{20}\putm{20}\putm{20}
\putm{20}\putp{20}\putm{20}\putm{-210}\addtocounter{nny}{17}
\putm{40}\putm{40}\putp{40}\putp{40}\putp{40}\putp{-210}\addtocounter{nny}{17}
\putm{20}\putp{20}\putp{20}\putp{20}
\putm{20}\putm{20}\putp{20}\putm{20}
\putm{20}\putm{20}\putp{20}\putm{-190}\addtocounter{nny}{17}
\putm{40}\putm{40}\putp{40}\putp{40}\putp{40}\putp{-230}\addtocounter{nny}{17}
\putp{20}\putm{20}\putp{20}\putp{20}
\putp{20}\putm{20}\putm{20}\putp{20}
\putm{20}\putm{20}\putm{20}\putp{-210}\addtocounter{nny}{17}
\putm{40}\putm{40}\putm{40}\putp{40}\putp{40}\putp{-210}\addtocounter{nny}{17}
\putp{20}\putp{20}\putm{20}\putp{20}
\putp{20}\putp{20}\putm{20}\putm{20}
\putp{20}\putm{20}\putm{20}\putm{-190}\addtocounter{nny}{17}
\putm{40}\putm{40}\putm{40}\putp{40}\putp{40}\putp{-230}\addtocounter{nny}{17}
\putp{20}\putp{20}\putp{20}\putm{20}
\putp{20}\putp{20}\putp{20}\putm{20}
\putm{20}\putp{20}\putm{20}\putm{-210}\addtocounter{nny}{17}
\putm{40}\putm{40}\putm{40}\putm{40}\putp{40}\putp{-210}\addtocounter{nny}{17}
\putm{20}\putp{20}\putp{20}\putp{20}
\putm{20}\putp{20}\putp{20}\putp{20}
\putm{20}\putm{20}\putp{20}\putm{-210}\addtocounter{nny}{17}
\end{picture}
\end{center}

\begin{center}
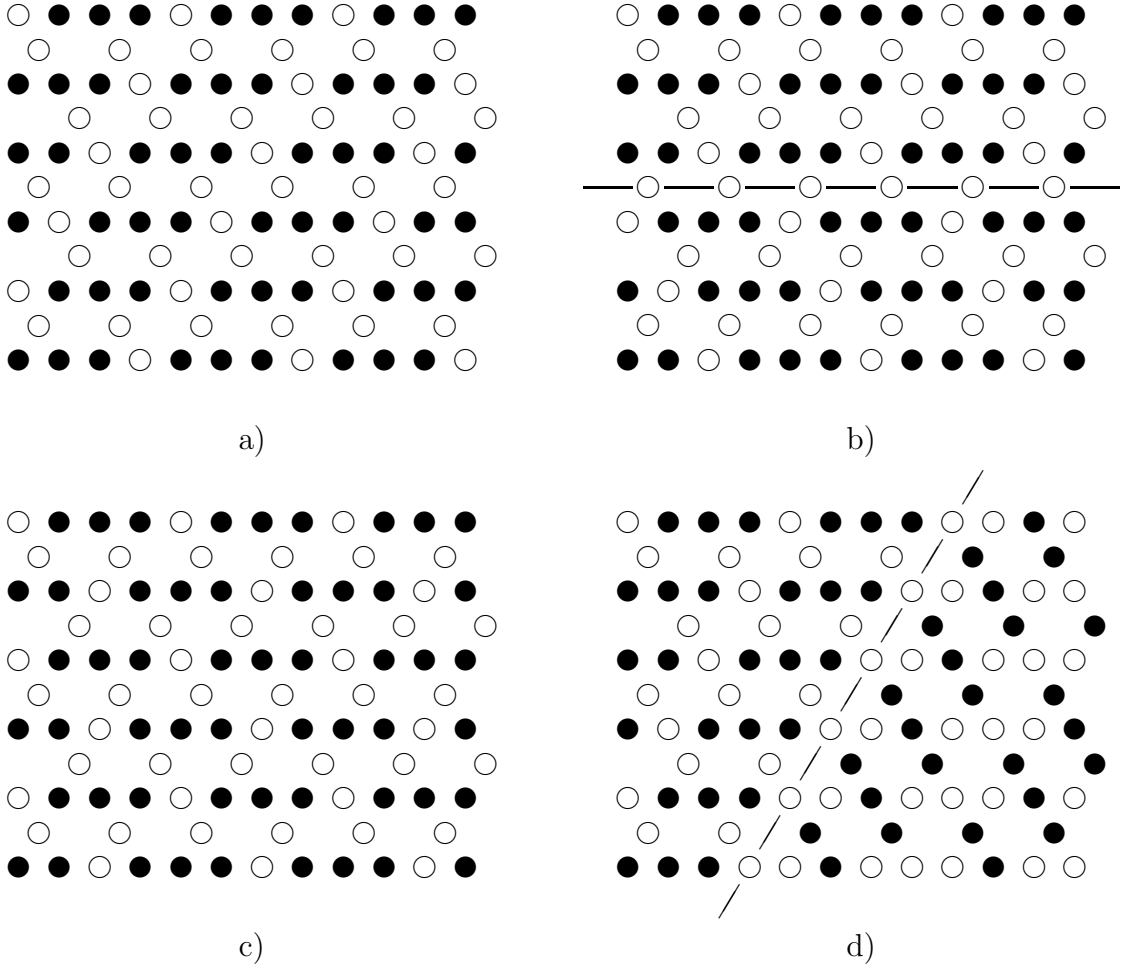
(the first part of Figure 2)
\end{center}

\newpage
\begin{center}
\begin{picture}(530,450)(0,-750)
\stc{000}{-470} \putl{e)}

\putm{20}\putp{20}\putp{20}\putm{20}
\putm{20}\putp{20}\putp{20}\putm{20}
\putm{20}\putp{20}\putp{20}\putm{-210}\addtocounter{nny}{17}
\putp{40}\putm{40}\putp{40}\putm{40}\putp{40}\putm{-210}\addtocounter{nny}{17}
\putm{20}\putm{20}\putp{20}\putp{20}
\putm{20}\putm{20}\putp{20}\putp{20}
\putm{20}\putm{20}\putp{20}\putp{-190}\addtocounter{nny}{17}
\putp{40}\putm{40}\putp{40}\putm{40}\putp{40}\putm{-230}\addtocounter{nny}{17}
\putp{20}\putm{20}\putm{20}\putp{20}
\putp{20}\putm{20}\putm{20}\putp{20}
\putp{20}\putm{20}\putm{20}\putp{-210}\addtocounter{nny}{17}
\putm{40}\putp{40}\putm{40}\putp{40}\putm{40}\putp{-210}\addtocounter{nny}{17}
\putp{20}\putp{20}\putm{20}\putm{20}
\putp{20}\putp{20}\putm{20}\putm{20}
\putp{20}\putp{20}\putm{20}\putm{-190}\addtocounter{nny}{17}
\putm{40}\putp{40}\putm{40}\putp{40}\putm{40}\putp{-230}\addtocounter{nny}{17}
\putm{20}\putp{20}\putp{20}\putm{20}
\putm{20}\putp{20}\putp{20}\putm{20}
\putm{20}\putp{20}\putp{20}\putm{-210}\addtocounter{nny}{17}
\putp{40}\putm{40}\putp{40}\putm{40}\putp{40}\putm{-210}\addtocounter{nny}{17}
\putm{20}\putm{20}\putp{20}\putp{20}
\putm{20}\putm{20}\putp{20}\putp{20}
\putm{20}\putm{20}\putp{20}\putp{-210}\addtocounter{nny}{17}

\stc{300}{-606} \putl{f)}
\addtocounter{nnx}{+5}\addtocounter{nny}{-25}
\multiput(\value{nnx},\value{nny})(20,34){4}{\line(3,5){10}}
\addtocounter{nnx}{-5}\addtocounter{nny}{+25}
\addtocounter{nnx}{+125}\addtocounter{nny}{-25}
\multiput(\value{nnx},\value{nny})(20,34){4}{\line(3,5){10}}
\addtocounter{nnx}{-125}\addtocounter{nny}{+25}
\addtocounter{nnx}{+45}\addtocounter{nny}{+247}
\multiput(\value{nnx},\value{nny})(20,34){3}{\line(3,5){10}}
\addtocounter{nnx}{-45}\addtocounter{nny}{-247}
\addtocounter{nnx}{+165}\addtocounter{nny}{+247}
\multiput(\value{nnx},\value{nny})(20,34){3}{\line(3,5){10}}
\addtocounter{nnx}{-165}\addtocounter{nny}{-247}
\addtocounter{nnx}{+75}\addtocounter{nny}{145}
\multiput(\value{nnx},\value{nny})(-20,34){2}{\line(-3,5){10}}
\addtocounter{nnx}{-75}\addtocounter{nny}{-145}
\addtocounter{nnx}{+195}\addtocounter{nny}{145}
\multiput(\value{nnx},\value{nny})(-20,34){2}{\line(-3,5){10}}
\addtocounter{nnx}{-195}\addtocounter{nny}{-145}
\addtocounter{nnx}{+90}\addtocounter{nny}{119}
\multiput(\value{nnx},\value{nny})(120,0){2}{\line(-3,5){8}}
\multiput(\value{nnx},\value{nny})(120,0){2}{\line(-3,-5){8}}
\addtocounter{nnx}{-90}\addtocounter{nny}{-119}
\addtocounter{nnx}{+30}\addtocounter{nny}{221}
\multiput(\value{nnx},\value{nny})(120,0){2}{\line(+3,5){8}}
\multiput(\value{nnx},\value{nny})(120,0){2}{\line(+3,-5){8}}
\addtocounter{nnx}{-30}\addtocounter{nny}{-221}

\putp{20}\putm{20}\putm{20}\putp{20}
\putm{20}\putm{20}\putm{20}\putp{20}
\putp{20}\putm{20}\putp{20}\putp{-210}\addtocounter{nny}{17}
\putm{40}\putp{40}\putp{40}\putp{40}\putm{40}\putm{-210}\addtocounter{nny}{17}
\putp{20}\putp{20}\putm{20}\putm{20}
\putp{20}\putm{20}\putm{20}\putm{20}
\putp{20}\putp{20}\putm{20}\putp{-190}\addtocounter{nny}{17}
\putm{40}\putp{40}\putp{40}\putp{40}\putm{40}\putm{-230}\addtocounter{nny}{17}
\putp{20}\putp{20}\putp{20}\putm{20}
\putm{20}\putp{20}\putm{20}\putm{20}
\putm{20}\putp{20}\putp{20}\putm{-210}\addtocounter{nny}{17}
\putm{40}\putm{40}\putp{40}\putp{40}\putp{40}\putm{-210}\addtocounter{nny}{17}
\putm{20}\putp{20}\putp{20}\putp{20}
\putm{20}\putm{20}\putp{20}\putm{20}
\putm{20}\putm{20}\putp{20}\putp{-190}\addtocounter{nny}{17}
\addtocounter{nnx}{-32}
\multiput(\value{nnx},\value{nny})(40,0){7}{\line(1,0){24}}
\addtocounter{nnx}{+32}
\putm{40}\putm{40}\putp{40}\putp{40}\putp{40}\putm{-230}\addtocounter{nny}{17}
\putp{20}\putp{20}\putm{20}\putp{20}
\putp{20}\putm{20}\putm{20}\putm{20}
\putp{20}\putm{20}\putm{20}\putp{-210}\addtocounter{nny}{17}
\putm{40}\putm{40}\putp{40}\putp{40}\putp{40}\putm{-210}\addtocounter{nny}{17}
\putp{20}\putm{20}\putp{20}\putp{20}
\putm{20}\putm{20}\putm{20}\putp{20}
\putm{20}\putm{20}\putp{20}\putp{-190}\addtocounter{nny}{17}
\putm{40}\putp{40}\putp{40}\putp{40}\putm{40}\putm{-230}\addtocounter{nny}{17}
\putm{20}\putp{20}\putp{20}\putm{20}
\putm{20}\putm{20}\putp{20}\putm{20}
\putm{20}\putp{20}\putp{20}\putp{-210}\addtocounter{nny}{17}
\addtocounter{nnx}{-32}
\multiput(\value{nnx},\value{nny})(40,0){7}{\line(1,0){24}}
\addtocounter{nnx}{+32}
\putm{40}\putp{40}\putp{40}\putp{40}\putm{40}\putm{-210}\addtocounter{nny}{17}
\putp{20}\putp{20}\putm{20}\putm{20}
\putp{20}\putm{20}\putm{20}\putm{20}
\putp{20}\putp{20}\putm{20}\putp{-190}\addtocounter{nny}{17}
\putm{40}\putp{40}\putp{40}\putp{40}\putm{40}\putm{-230}\addtocounter{nny}{17}
\putp{20}\putp{20}\putp{20}\putm{20}
\putm{20}\putp{20}\putm{20}\putm{20}
\putm{20}\putp{20}\putp{20}\putm{-210}\addtocounter{nny}{17}
\putm{40}\putm{40}\putp{40}\putp{40}\putp{40}\putm{-210}\addtocounter{nny}{17}
\putm{20}\putp{20}\putp{20}\putp{20}
\putm{20}\putm{20}\putp{20}\putm{20}
\putm{20}\putm{20}\putp{20}\putp{-210}\addtocounter{nny}{17}

\stc{000}{-720} \putl{g)}

\putp{20}\putm{20}\putm{20}\putp{20}
\putp{20}\putm{20}\putm{20}\putp{20}
\putp{20}\putm{20}\putm{20}\putp{-210}\addtocounter{nny}{17}
\putp{40}\putm{40}\putp{40}\putm{40}\putp{40}\putm{-210}\addtocounter{nny}{17}
\putm{20}\putm{20}\putp{20}\putp{20}
\putm{20}\putm{20}\putp{20}\putp{20}
\putm{20}\putm{20}\putp{20}\putp{-190}\addtocounter{nny}{17}
\putp{40}\putm{40}\putp{40}\putm{40}\putp{40}\putm{-230}\addtocounter{nny}{17}
\putp{20}\putm{20}\putm{20}\putp{20}
\putp{20}\putm{20}\putm{20}\putp{20}
\putp{20}\putm{20}\putm{20}\putp{-210}\addtocounter{nny}{17}
\putp{40}\putm{40}\putp{40}\putm{40}\putp{40}\putm{-210}\addtocounter{nny}{17}
\putm{20}\putm{20}\putp{20}\putp{20}
\putm{20}\putm{20}\putp{20}\putp{20}
\putm{20}\putm{20}\putp{20}\putp{-190}\addtocounter{nny}{17}
\putp{40}\putm{40}\putp{40}\putm{40}\putp{40}\putm{-230}\addtocounter{nny}{17}
\putp{20}\putm{20}\putm{20}\putp{20}
\putp{20}\putm{20}\putm{20}\putp{20}
\putp{20}\putm{20}\putm{20}\putp{-210}\addtocounter{nny}{17}
\putp{40}\putm{40}\putp{40}\putm{40}\putp{40}\putm{-210}\addtocounter{nny}{17}
\putm{20}\putm{20}\putp{20}\putp{20}
\putm{20}\putm{20}\putp{20}\putp{20}
\putm{20}\putm{20}\putp{20}\putp{-190}\addtocounter{nny}{17}
\end{picture}
\end{center}

\begin{figure}
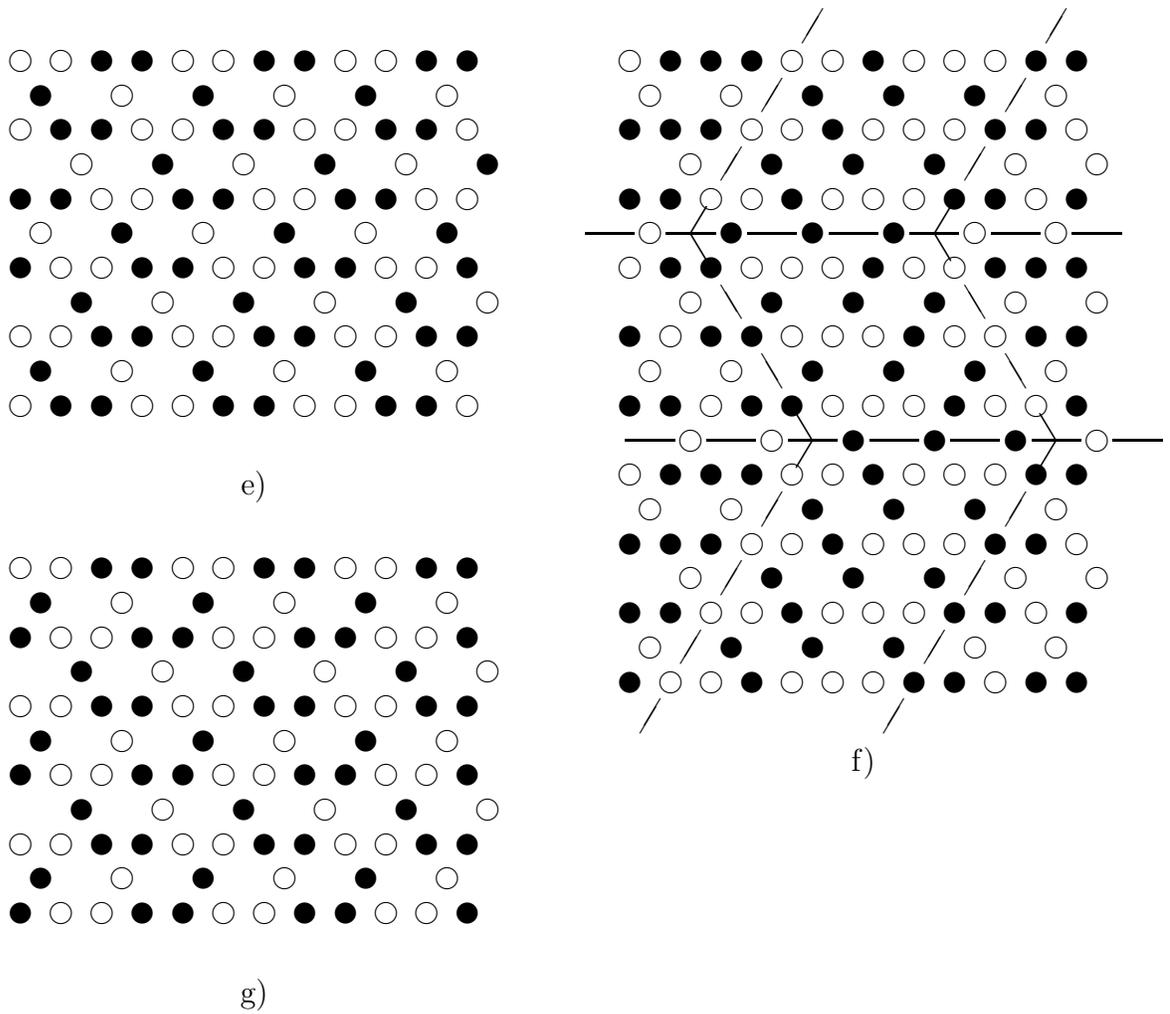

\caption{Filled (empty) circles designate positive (negative)
half-vortices. Structures a), c), e) and g) are periodic, whereas
b), d) and f) include zero-energy domain walls separating
different periodic states. All states shown have the same
energy.\\}
\end{figure}

\setlength{\unitlength}{0.3mm}
\begin{center}
\begin{picture}(480,200)(0,-40)
\thicklines
\newcommand{\putll}[1]{
\addtocounter{nnx}{+90}\addtocounter{nny}{-30}
\put(\value{nnx},\value{nny}){\makebox(0,0)[cc]{#1}}
\addtocounter{nnx}{-90}\addtocounter{nny}{+30}}
\newcommand{\pzu}[2]{
\addtocounter{nnx}{#1}\addtocounter{nny}{#2}
\put(\value{nnx},\value{nny}){\vector(0,1){40}}
\addtocounter{nnx}{-#1}\addtocounter{nny}{-#2}}
\newcommand{\pzd}[2]{
\addtocounter{nnx}{#1}\addtocounter{nny}{#2}
\put(\value{nnx},\value{nny}){\vector(0,-1){40}}
\addtocounter{nnx}{-#1}\addtocounter{nny}{-#2}}
\newcommand{\pzru}[2]{
\addtocounter{nnx}{#1}\addtocounter{nny}{#2}
\put(\value{nnx},\value{nny}){\vector(3,2){30}}
\addtocounter{nnx}{-#1}\addtocounter{nny}{-#2}}
\newcommand{\pzrd}[2]{
\addtocounter{nnx}{#1}\addtocounter{nny}{#2}
\put(\value{nnx},\value{nny}){\vector(3,-2){30}}
\addtocounter{nnx}{-#1}\addtocounter{nny}{-#2}}
\newcommand{\pzlu}[2]{
\addtocounter{nnx}{#1}\addtocounter{nny}{#2}
\put(\value{nnx},\value{nny}){\vector(-3,+2){30}}
\addtocounter{nnx}{-#1}\addtocounter{nny}{-#2}}
\newcommand{\pzld}[2]{
\addtocounter{nnx}{#1}\addtocounter{nny}{#2}
\put(\value{nnx},\value{nny}){\vector(-3,-2){30}}
\addtocounter{nnx}{-#1}\addtocounter{nny}{-#2}}
\newcommand{\pdu}[2]{
\addtocounter{nnx}{#1}\addtocounter{nny}{#2}
\put(\value{nnx},\value{nny}){\vector(0,1){40}}
\put(\value{nnx},\value{nny}){\vector(0,1){34}}
\addtocounter{nnx}{-#1}\addtocounter{nny}{-#2}}
\newcommand{\pdd}[2]{
\addtocounter{nnx}{#1}\addtocounter{nny}{#2}
\put(\value{nnx},\value{nny}){\vector(0,-1){40}}
\put(\value{nnx},\value{nny}){\vector(0,-1){34}}
\addtocounter{nnx}{-#1}\addtocounter{nny}{-#2}}
\newcommand{\pdru}[2]{
\addtocounter{nnx}{#1}\addtocounter{nny}{#2}
\put(\value{nnx},\value{nny}){\vector(3,2){30}}
\put(\value{nnx},\value{nny}){\vector(3,2){25}}
\addtocounter{nnx}{-#1}\addtocounter{nny}{-#2}}
\newcommand{\pdrd}[2]{
\addtocounter{nnx}{#1}\addtocounter{nny}{#2}
\put(\value{nnx},\value{nny}){\vector(3,-2){30}}
\put(\value{nnx},\value{nny}){\vector(3,-2){25}}
\addtocounter{nnx}{-#1}\addtocounter{nny}{-#2}}
\newcommand{\pdlu}[2]{
\addtocounter{nnx}{#1}\addtocounter{nny}{#2}
\put(\value{nnx},\value{nny}){\vector(-3,+2){30}}
\put(\value{nnx},\value{nny}){\vector(-3,+2){25}}
\addtocounter{nnx}{-#1}\addtocounter{nny}{-#2}}
\newcommand{\pdld}[2]{
\addtocounter{nnx}{#1}\addtocounter{nny}{#2}
\put(\value{nnx},\value{nny}){\vector(-3,-2){30}}
\put(\value{nnx},\value{nny}){\vector(-3,-2){25}}
\addtocounter{nnx}{-#1}\addtocounter{nny}{-#2}}
\newcommand{\ptu}[2]{
\addtocounter{nnx}{#1}\addtocounter{nny}{#2}
\put(\value{nnx},\value{nny}){\vector(0,1){40}}
\put(\value{nnx},\value{nny}){\vector(0,1){34}}
\put(\value{nnx},\value{nny}){\vector(0,1){28}}
\addtocounter{nnx}{-#1}\addtocounter{nny}{-#2}}
\newcommand{\ptd}[2]{
\addtocounter{nnx}{#1}\addtocounter{nny}{#2}
\put(\value{nnx},\value{nny}){\vector(0,-1){40}}
\put(\value{nnx},\value{nny}){\vector(0,-1){34}}
\put(\value{nnx},\value{nny}){\vector(0,-1){28}}
\addtocounter{nnx}{-#1}\addtocounter{nny}{-#2}}
\newcommand{\ptru}[2]{
\addtocounter{nnx}{#1}\addtocounter{nny}{#2}
\put(\value{nnx},\value{nny}){\vector(3,2){30}}
\put(\value{nnx},\value{nny}){\vector(3,2){25}}
\put(\value{nnx},\value{nny}){\vector(3,2){20}}
\addtocounter{nnx}{-#1}\addtocounter{nny}{-#2}}
\newcommand{\ptrd}[2]{
\addtocounter{nnx}{#1}\addtocounter{nny}{#2}
\put(\value{nnx},\value{nny}){\vector(3,-2){30}}
\put(\value{nnx},\value{nny}){\vector(3,-2){25}}
\put(\value{nnx},\value{nny}){\vector(3,-2){20}}
\addtocounter{nnx}{-#1}\addtocounter{nny}{-#2}}
\newcommand{\ptlu}[2]{
\addtocounter{nnx}{#1}\addtocounter{nny}{#2}
\put(\value{nnx},\value{nny}){\vector(-3,2){30}}
\put(\value{nnx},\value{nny}){\vector(-3,2){25}}
\put(\value{nnx},\value{nny}){\vector(-3,2){20}}
\addtocounter{nnx}{-#1}\addtocounter{nny}{-#2}}
\newcommand{\ptld}[2]{
\addtocounter{nnx}{#1}\addtocounter{nny}{#2}
\put(\value{nnx},\value{nny}){\vector(-3,-2){30}}
\put(\value{nnx},\value{nny}){\vector(-3,-2){25}}
\put(\value{nnx},\value{nny}){\vector(-3,-2){20}}
\addtocounter{nnx}{-#1}\addtocounter{nny}{-#2}}
\newcommand{\ppp}[2]{
\addtocounter{nnx}{#1}\addtocounter{nny}{#2}
\put(\value{nnx},\value{nny}){\makebox(0,0)[cc]{$+$}}
\addtocounter{nnx}{-#1}\addtocounter{nny}{-#2}}
\newcommand{\ppm}[2]{
\addtocounter{nnx}{#1}\addtocounter{nny}{#2}
\put(\value{nnx},\value{nny}){\makebox(0,0)[cc]{$-$}}
\addtocounter{nnx}{-#1}\addtocounter{nny}{-#2}}
\newcommand{\pppp}[2]{
\addtocounter{nnx}{#1}\addtocounter{nny}{#2}
\put(\value{nnx},\value{nny}){\makebox(0,0)[cc]{$[+]$}}
\addtocounter{nnx}{-#1}\addtocounter{nny}{-#2}}
\newcommand{\pppm}[2]{
\addtocounter{nnx}{#1}\addtocounter{nny}{#2}
\put(\value{nnx},\value{nny}){\makebox(0,0)[cc]{$[-]$}}
\addtocounter{nnx}{-#1}\addtocounter{nny}{-#2}}

\stc{0}{0} \putll{a)}

\pzru{60}{0} \pdlu{60}{0} \pdlu{120}{0} \ptru{30}{20}
\ptu{090}{20} \ptru{150}{20} \ptld{150}{20} \pdrd{0}{40}
\pzu{60}{40} \pdrd{60}{40} \pzu{180}{40} \pzld{120}{40}
\pdrd{120}{40} \pzd{30}{60} \pdlu{30}{60} \pzru{090}{60}
\pdlu{090}{60} \pzd{150}{60} \pdlu{150}{60} \ptu{0}{080}
\ptd{0}{080} \ptld{60}{080} \ptru{60}{080} \ptu{120}{080}
\ptd{120}{080} \ptld{180}{080} \pzld{30}{100} \pdrd{30}{100}
\pzu{090}{100} \pdrd{090}{100} \pzld{150}{100} \pdrd{150}{100}
\pzru{0}{120} \pzd{60}{120} \pdlu{60}{120} \pzru{120}{120}
\pdlu{120}{120} \pzd{180}{120} \pdlu{180}{120} \ptd{30}{140}
\ptld{090}{140} \ptru{090}{140} \ptd{150}{140} \pzld{60}{160}
\pdrd{60}{160} \pdrd{120}{160}

\ppm{60}{20} \ppm{120}{20} \ppp{15}{50} \pppp{40}{50}
\ppp{075}{50} \pppm{105}{50} \ppp{135}{50} \pppp{165}{50}
\ppm{30}{080} \ppm{090}{080} \ppm{150}{080} \pppm{15}{110}
\ppp{45}{110} \pppp{075}{110} \ppp{105}{110} \pppm{135}{110}
\ppp{165}{110} \ppm{60}{140} \ppm{120}{140}

\stc{210}{0} \addtocounter{nnx}{-60}\putll{b)}
\addtocounter{nnx}{60} \ptu{30}{50} \pdlu{30}{90} \pzru{30}{90}

\stc{300}{0} \putll{c)} \ptlu{60}{0} \pdru{090}{20} \pdld{090}{20}
\pdru{30}{20} \ptd{090}{60} \pdru{150}{20} \pdld{150}{20}
\ptrd{0}{40} \ptu{60}{40} \pzlu{090}{20} \pzrd{090}{20}
\ptrd{120}{40} \ptu{180}{40} \pzu{30}{20} \ptlu{30}{60}
\pdld{120}{080} \pzrd{60}{080} \pzu{150}{20} \ptlu{150}{60}
\ptu{0}{080} \pzd{0}{080} \pdld{60}{080} \ptru{60}{080}
\ptu{120}{080} \pzd{120}{080} \pdld{180}{080} \pzld{30}{100}
\pdrd{30}{100} \pzu{090}{100} \pdrd{090}{100} \pzld{150}{100}
\pdrd{150}{100} \pzru{0}{120} \pzd{60}{120} \pdlu{60}{120}
\pzru{120}{120} \pdlu{120}{120} \pzd{180}{120} \pdlu{180}{120}
\ptd{30}{140} \ptld{090}{140} \ptru{090}{140} \ptd{150}{140}
\pzld{60}{160} \pdrd{60}{160} \pdrd{120}{160}

\ppm{60}{20} \ppm{120}{20} \pppp{15}{50} \ppp{45}{50}
\pppm{075}{50} \ppp{105}{50} \pppp{135}{50} \ppp{165}{50}
\ppm{30}{080} \ppm{090}{080} \ppm{150}{080} \pppm{15}{110}
\ppp{45}{110} \pppp{075}{110} \ppp{105}{110} \pppm{135}{110}
\ppp{165}{110} \ppm{60}{140} \ppm{120}{140}
\end{picture}
\end{center}

\begin{figure}
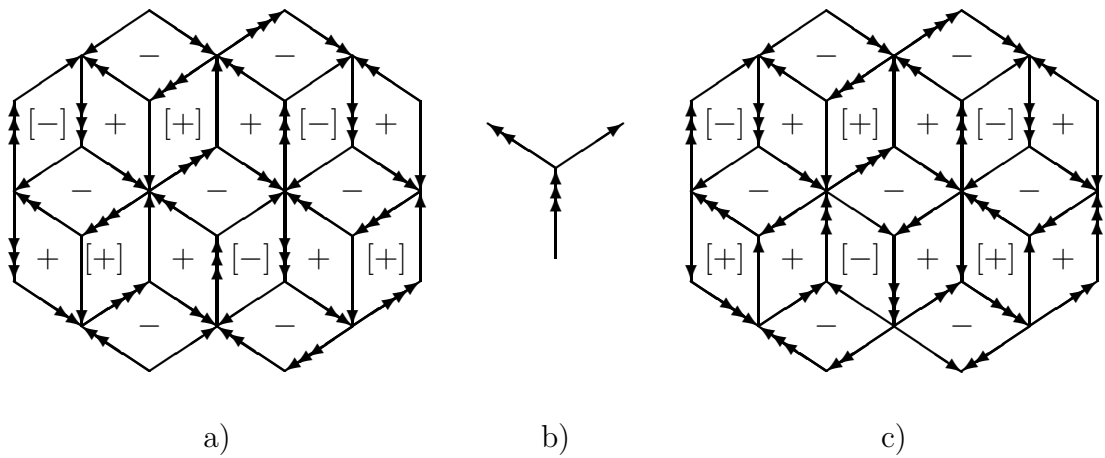

\caption{a) Phase representation of the periodic state shown in
Fig. 2a; b) elementary pattern, repetition of which allows to
construct this state; c) phase representation of a zero-energy
domain wall (shown in Fig. 2b). Three types of arrows correspond
to three different values of $\theta_{ij}$.}
\end{figure}
\end{document}